\documentstyle[epsf,graphics]{mn}

\newcommand{\Msolar}{\mbox{\,$\rm M_{\odot}$}}        
\newcommand{\asec}{^{\prime\prime}}
\title[The Evolution of 3CR Radio Galaxies from z=1]{ The Evolution of 3CR Radio Galaxies from z=1}
\author[R.J. McLure and J.S. Dunlop]
{R.J. McLure$^{1,2}$ and J.S. Dunlop$^{1}$ 
\\
$^{1}$Institute for Astronomy, University of Edinburgh, Blackford Hill,
Edinburgh, EH9~3HJ\\
$^{2}$Nuclear and Astrophysics Laboratory, University of Oxford, Keble
Road, Oxford, OX1 3RH}

\date{Submitted for publication in MNRAS}
\begin{document}
\maketitle

\begin{abstract}
We present the results of a comprehensive re-analysis of the
 images of a virtually complete sample of 28 powerful 3CR radio
 galaxies with redshifts $0.6<z<1.8$ from the HST archive. Using a 
two-dimensional modelling technique we have derived scalelengths and 
absolute magnitudes for a total of 16 3CR galaxies with a median 
redshift of $z=0.8$. Our results confirm the basic conclusions of
 Best, Longair \& R\"{o}ttgering (1997, 1998) in that we also find $z=1$ 3CR
 galaxies to be massive, well-evolved ellipticals, whose infrared
 emission is dominated by starlight. However,  we in fact find 
that the scalelength
 distribution of 3CR galaxies at $z \simeq 1$ is completely indistinguishable
 from that derived for their low-redshift counterparts from our own
recently-completed HST study of AGN hosts at $z \simeq 0.2$. There is
 thus no evidence that 
3CR radio galaxies at $z \simeq 1$ are dynamically different from 3CR galaxies 
at low redshift. Moreover, for a 10-object sub-sample we have
 determined the galaxy parameters with sufficient accuracy to
 demonstrate, for the first time, that the $z \simeq 1$ 3CR galaxies
 follow a Kormendy relation which is indistinguishable from that
 displayed by low-redshift ellipticals if one allows for purely
 passive evolution. The implied rather modest level of passive
 evolution since $z \simeq 1$ is consistent with that predicted from 
spectrophotometric models provided one assumes a high formation
 redshift ($z \ge 4$) within a low-density Universe. We conclude that 
there is no convincing evidence for significant dynamical evolution
 among 3CR galaxies in the redshift interval $0<z<1$, and that simple
 passive evolution remains an acceptable interpretation of the $K-z$ 
relation for powerful radio galaxies.
\end{abstract}
\section{Introduction}
\label{background}
\begin{figure}
\setlength{\epsfxsize}{0.35\textwidth}
\centerline{\epsfbox{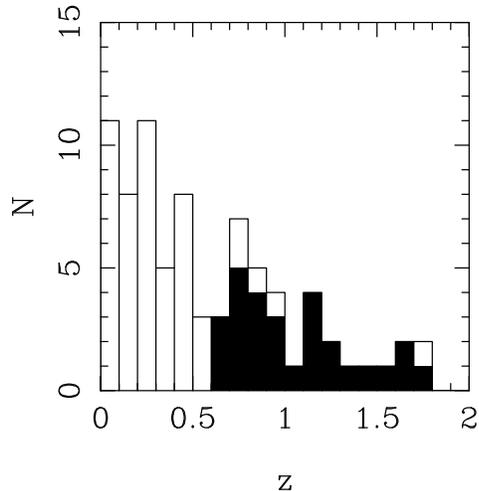}}
\caption{The redshift distribution of the complete 79 object FR{\sc II} 3CR subsample 
(Laing {\it et al.} 1983). The shaded objects are those included in the Best {\it et al.} 
sample. Reproduced from Best {\it et al.} 1997.}
\label{3csample}
\end{figure}

It has been known since the early 1980's that the hosts of powerful radio galaxies display a tight 
relation between the $K$-band magnitude and redshift (Lilly\,\&\,Longair 1984). In recent years it 
has been shown that, at least out to $z = 1$, essentially the same relation is followed by both the 
less powerful Parkes Selected Regions (PSR) and 6C radio galaxies 
(Dunlop {\it et al.} 1989; Eales {\it et al.} 1997) and by X-ray
selected brightest cluster galaxies (Collins \& Mann 1997). When the $K$-band magnitudes of the 
3CR galaxies are corrected for the expected effects of passive evolution of their stellar 
populations they appear to represent rather good standard candles from the present day out to 
redshift one and greater. Consequently, for many years it was widely accepted that the tightness
of the $K-z$ relation for the 3CR radio galaxies could be most naturally explained by the most 
powerful radio galaxies having a rather well-defined mass, being formed at $z \gg 1$, and evolving 
basically passively thereafter.

However, this long-held view has recently been challenged in a series of papers by Best, Longair \& 
R\"{o}ttgering (1997,1998), hereafter BLR. Between 1994--1996 BLR undertook an extensive study of a 
virtually complete sample of $28$ powerful FR{\sc II} 3CR galaxies in the redshift range $0.6<z<1.8$ 
(see Fig \ref{3csample}). BLR obtained V and I--band HST images, J and K--band UKIRT images, and 
complimentary radio observations at 8.4 GHz with the VLA.

BLR made use of the four broad--band images they obtained for each of their sources to perform 
spectral synthesis fitting.  The four broad--band fluxes where fitted by a simple two-component 
model consisting of an old stellar population and a power-law contribution which represented any 
possible aligned component. The stellar population SED's were constructed from the models of
Bruzual \& Charlot (1993), and assumed a 1-Gyr burst of star formation at  $z_{for}=10$, with
 the stars evolving passively from that point onwards to the redshift of the particular 3CR galaxy. 
The assumed form of any possible nuclear/aligned emission was a power-law of the form 
$f_{\nu}\,\,\alpha\,\,\nu^{-0.4}$ which was used as a compromise between the likely contributions of 
quasar/starburst/dust scattered emission.  The results of this fitting procedure (BLR 1998) show that 
the broad--band fluxes from these 3CR galaxies are fully consistent with that expected from an 
old passively evolving stellar population, as required in the traditional interpretation of the $K-z$ 
relation.

However, the results of the determination of the elliptical galaxy half-light radii $(r_{e})$ 
performed by BLR, combined with a summary of the literature concerning the environments of 3CR 
galaxies at high and low redshift, led BLR to challenge the traditional view of the $K-z$ relation.  
Using the method of fitting de Vaucouleurs templates to azimuthally averaged luminosity profiles, 
BLR produced best--fit scalelengths for 19 of their 28 objects (excluding 3C41 and 3C22 which have 
a large nuclear contribution (Leyshon \& Eales 1998) not dealt with in the BLR fitting scheme).  
The results of this profile fitting revealed a mean scalelength of $r_{e}=14.7\pm1.3$ kpc 
($\Omega_{0}=1, H_{0}=50$).  As pointed out by BLR this is significantly larger than the 
$r_{e}=8.2\pm1$ kpc found to be typical of low redshift ellipticals by Schombert (1987), and 
only around a factor of two smaller than the average scalelength of low-$z$ brightest cluster 
galaxies (BCG), $r_{e}=32.7\pm1.1$ kpc, found by the same author. Combining this result with the 
$L\,\alpha \,\,r_{e}^{0.7}$ relation found by Kormendy (1977), and confirmed by our HST AGN host 
galaxy study (Dunlop {\it et al} 2000), BLR argue that even at $z=1$ the hosts of 3CR radio galaxies 
are highly evolved massive systems, possibly larger by a factor of two than their low-$z$ counterparts.

Combining the apparent similarity of the $z=1$ 3CR galaxies to BCG at
the same redshift, with results from the 
literature concerning the environments of high redshift radio 
galaxies (see BLR 1998 for a review), BLR concluded that the passive 
appearance of the $K-z$ relation is a cosmic conspiracy. The general 
picture given by existing studies is that the environments of $z\simeq1$ 3CR's are consistent with 
moderately rich clusters; galaxy--galaxy cross-correlation function measures are consistent with 
Abell class 0, or richer (Hill \& Lilly 1991), while multi-colour imaging shows many companion 
objects consistent with passively evolving coeval galaxies (Dickinson 1997). As pointed out by BLR, 
the same is not found in existing studies of the environments of low redshift 3CR's 
(Prestage and Peacock 1988), where it has been reported that these objects reside in 
fairly low-density environments and have smaller characteristic sizes and luminosities than their 
high-redshift counterparts (Lilly and Prestage 1987). 

On the basis of the apparent redshift dependence of 3CR cluster environments
BLR argued that the 3CR galaxies at $z=0$ and $z=1$ cannot be linked by a simple ``closed-box'' 
passive evolution scheme.  In the picture favoured by BLR, the $K-z$ relation can be explained by 
the two 3CR galaxy samples having different dynamical evolutionary histories, but being observed at 
the points in their respective histories where they both contain a characteristic mass of stars; 
a few times $10^{11}\Msolar$.  The $z\simeq1$ 3CR galaxies would reach this mass first due to their 
location in a larger density peak producing strong merging between $z=3\rightarrow1$.  
The 3CR galaxies at low-$z$ would reach the characteristic mass later through merging within their 
weaker cluster environment between $z=1$ and the present day, as is expected in many hierarchical 
galaxy formation models (Kauffmann 1999).  The 3CR galaxies at $z=1$ would obviously also grow 
through merger activity in this redshift range and would become the radio dormant BCG we see today 
at low redshift, radio activity having ceased presumably due to the lack of available gas to feed 
their central black-hole. Although this picture is both consistent and
appealing, it is dependent on the acceptance of
environmental studies which now date back $15-20$ years, together with
the finding that 3CR galaxies at $z\simeq1$ have sizes comparable to BCG at
that epoch, something which is not found to be true for low-$z$ 3CR
galaxies.

It was a careful examination of the methods used by BLR in determining their scalelengths which 
provided the original motivation for this paper.  As mentioned above, scalelengths were presented by 
BLR for 19 of their 28 object sample (the images of the excluded 9 objects having poor signal-to-noise or suffering from extreme aligned emission).  However, only 6 of these 19 scalelengths where actually 
determined from the high resolution HST imaging. The reason for this is that the one-dimensional 
profile fitting technique employed was unable to cope with the masking of significant contamination 
by companion objects or aligned emission.  The majority of the scalelength information presented by 
BLR was in fact derived from the analysis of their $K$-band imaging data obtained at 
UKIRT.  As reported by BLR, this imaging was obtained in $1\asec$ seeing conditions with typically 
54 minutes of on-source integration which, considering the redshift range of the objects, leaves a 
relatively small amount of data which has both good signal-to-noise and is free from substantial 
seeing effects. The use of an analytical gausian function to represent what has proven to be the 
extremely complicated IRCAM3 PSF (McLure {\it et al} 1999b) casts further doubt on the quoted 
scalelength values.
\begin{figure}
\setlength{\epsfxsize}{0.3\textwidth}
\centerline{\epsfbox{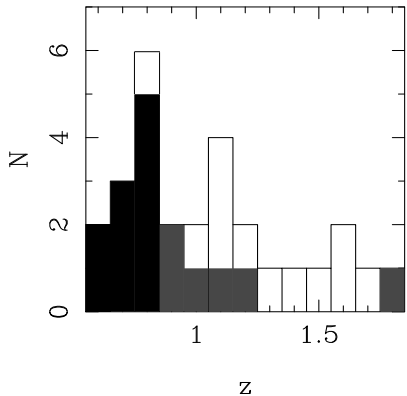}}
\caption{Histogram showing the redshift distribution of the virtually complete Best {\it et al.} sample. Objects shaded grey and black are objects which have been succesfully modelled during this re-analysis. Objects shaded in black are members of the 10-o
bject sub-sample.}
\label{subsample}
\end{figure}
Experience gained from the comparison of our modelling results of $z\simeq0.2$  AGN host galaxies 
(McLure {\it et al} 1999a, McLure {\it et al} 1999b) with those previously obtained from typically 
$1\asec$ $K$-band imaging (Dunlop {\it et al.} 1993, Taylor {\it et al.} 1996) has demonstrated to us
that the twin benefits of high resolution and a temporally stable PSF tend to combine to revise 
the best-fitting galaxy scalelength downwards.  The main reason for this systematic trend appears 
to be that the high spatial resolution provided by HST allows the reliable identification of separate 
companion objects and their subsequent masking from the modelling process, preventing the inclusion of extra flux from biasing the scalelength determination to higher values. This effect is obviously only 
going to be strengthened at redshifts of $z\sim1$ where the angular separation between companion 
objects and the target can easily be of the same order as the typical seeing experienced during the 
BLR UKIRT observations.

Given these considerations it was felt worthwhile to re-examine the publically available HST data to 
determine whether the use of the full two-dimensional analysis utilised in the $z\simeq0.2$ quasar host galaxy programme would produce significantly different scalelength figures from those published by BLR.

\section{Data Reduction}
\begin{table*}
\begin{tabular}{lclcccccc}
\hline
\hline
Source&z&Filter&Exp Time/s&$r_{1/2}$/$\asec$&$r_{1/2}$/kpc&$\mu_{1/2}$&$I_{c}$&$L_{nuc}$/$L_{host}$\\
\hline

3C22&0.938&F814W&1400&0.7&5.9&22.3&19.8&0.306\\

3C34$^{*}$&0.690&F785LP&1700&3.0&23.8&24.2&18.8&0.002\\

3C41$^{*}$&0.795&F785LP&1700&0.8&6.3&22.2&19.7&0.170\\

3C49$^{*}$&0.621&F814W&1400&2.0 &15.0&23.0&18.9&0.000\\

3C65&1.176&F814W&1760&1.2&9.9&24.5&20.9&0.088\\

3C217&0.897&F814W&1700&0.7&5.8&22.0&20.1&0.007\\

3C226$^{*}$&0.820&F785LP&1700&1.0&8.5&22.6&19.4&0.050\\

3c239&1.781&F814W&2200&1.3& 10.8 &23.5&19.7&0.041\\

3C247$^{*}$&0.749&F814W&2400&3.1&25.2&24.1&18.6&0.025\\

3c252&1.105&F814W&1700&0.9&7.3&23.2&20.5&0.000\\

3C277.2$^{*}$&0.766&F814W&2400&0.9&7.6&22.5&19.7&0.005\\

3C289&0.967&F814W&1800&2.1&18.0& 24.9&20.0&0.063\\

3C337$^{*}$&0.635&F814W&1400&0.8&6.4&22.5&19.8&0.055\\

3C340$^{*}$&0.775&F785LP&1700&0.5&4.3&21.3&19.7&0.000\\

3C352$^{*}$&0.806&F814W&1800&2.2&18.1&24.7&19.6&0.024\\

3C441$^{*}$&0.708&F785LP&1700&1.4&10.9&22.7&19.1&0.023\\
\hline
\hline
\end{tabular}
\caption{The results of the two-dimensional modelling of the
16-objects from the Best {\it et al.} $z\simeq1$ 3CR sample which did
not suffer from excessive aligned emission. Listed in column 5 are the
fitted scalelengths in arcseconds, with the corresponding figure in
kpc for $\Omega_{0}=1, H_{0}=50$ listed in column 6. The
surface-brightness at the fitted scalelength in Cousins I-magnitudes
arcsec$^{-2}$ is given in column 7. Column 8 lists the integrated
apparent I--band magnitude of the best-fit host galaxy with column 9 
giving the ratio of the integrated luminosity of the best-fit nuclear 
component and host galaxy. Objects labelled with a $^{*}$ are members 
of the 10-object sub-sample.}
\label{bigtab}
\end{table*}

All the HST images taken of the 3CR galaxies in the 28-object BLR sample were obtained from the 
HST archive facility\footnote{http://archive.stsci.edu}.  A detailed list of redshifts, filters and 
exposure times can be found in Table 1 of BLR (1997). An investigation of the shorter-wavelength 
exposures of each object (mainly F555W, F622W) confirmed that they were either of insufficient 
signal-to-noise to be useful, or dominated by emission aligned with the radio axis. Of the 
28 objects in the sample, preliminary analysis of the $I$-band images revealed a total of 16 
which could be successfully modelled with our two-dimensional technique.  The observational 
parameters of these 16 objects are detailed in Table \ref{bigtab} and illustrated as part of the 
full sample in Fig \ref{subsample} .

For each object there are two {\sc cr-split} $I$-band exposures of
unequal length available, together giving a typical exposure time of 
$\simeq1800$ seconds. The initial processing of the 
images, flat-fielding and bias removal, was carried out by the 
standard HST pipeline.  The two exposures were then combined using 
the {\sc iraf} task {\sc crrej}, which successfully 
removes cosmic ray events using a sophisticated sigma-clipping
algorithm.  The next step in the reduction process was the fitting of
a plane to the image with the 3CR galaxy masked out, to 
accurately determine the sky background while allowing for any
residual flat-fielding gradients that may have been present.  The
final step was the production of a two-dimensional mask for each source 
which eliminated any companion objects, or aligned emission, from the 
model-fitting process, as well as any regions of the image that could 
have been biased by scattered light from nearby bright stars.  

In the vast majority of cases the high resolution provided by the HST
allowed the identification, and subsequent masking, of companion objects
and diffuse aligned emission in a straightforward fashion. For a small
number of sources (3C252 \&\ 3C239) the question of which areas of
the images should be masked was more ambiguous, due to the HST images 
revealing double nuclei which are not apparent in near-infrared images (BLR
1997). As mentioned by BLR, the detection of
two nuclear components in these sources suggests either that they are
the products of recent mergers, or alternatively, contain a central
dust lane. Given the undisturbed appearance of the K--band images of
these sources (BLR 1997), it was decided that no attempt would be made
to mask the central regions of the HST images before modelling. 

None of the objects comprising the 10-object sub-sample (see Section
\ref{ress}) appear to be highly disturbed in the HST images, and
consequently return model fits which are largely insensitive to the
details of aligned-emission masking. The advantage of being able to
mask areas of the HST images, while still running the model fitting 
in two-dimensions, has allowed us to take greater advantage of the HST 
data than was possible with the one-dimensional analysis techniques 
employed by BLR. 

\subsection{Empirical PSF Determination}

Simple radial surface-brightness plots for the 16 objects chosen for modelling reveals that there is 
sufficient signal-to-noise to allow fitting to a typical radius of $\simeq4\asec$. It is therefore 
necessary to have a point spread function (PSF) to convolve our model galaxies with which has good 
signal-to-noise out to at least this radius.  Due to the synthetic PSF's produced by the {\sc tinytim} software package (Krist 1995) being unable to reproduce the WFPC2's scattered light halo outside a 
radius of $\simeq 1.5\asec$, it is clear that an empirical PSF is required for the modelling of these 
data.  A glance at Table \ref{bigtab} reveals that all of the exposures utilised here were imaged 
through either the F785LP or F814W filters.  This required the acquisition of two relatively deep 
PSF's imaged with the correct filter/chip combination, which were also located close to the average 
chip position of the 3CR galaxies, in order to avoid the noticeable positional variation of the WFPC2 
PSF. An interrogation of the HST PSF search tool \footnote{http://www.stsci.edu/instruments/wfpc2/} 
produced disappointing results, with all available PSF's either too faint or too small in angular 
extent.

Fortunately, two suitable stars were present on the exposures of 3C41 (F785LP) and 3C239 (F814W).  Due to the need for sufficient depth in the PSF wings, both of these stars had saturated cores in even the shortest exposures.  To overcome this problem a modified version of the PSF re-sampling technique 
described in McLure {\it et al} 1999b was implemented.  Both PSF's were only saturated within a radius of $\le0.3\asec$ of their core, well inside the radius where {\sc tinytim} can accurately reproduce the empirical PSF. Making use of this, the two PSF's
 had their core replaced with the equivalent 
{\sc tinytim} model, the relative scaling being determined by matching the flux in an annulus between 
$0.4\asec<r<0.7\asec$. Another advantage of this approach is that due to {\sc tinytim's} ability to 
produce model PSF's at up to 50 times oversampling, sub-pixel centring of the PSF can be matched to 
that of the galaxies to an accuracy of $\leq 0.005\asec$. Although not as crucial as with the quasar 
host work it is still important to overcome the severe undersampling of WFPC2, especially as several 
of the 3CR galaxies modelled proved to have a substantial point-source contribution.
\section{Modelling}
\begin{table}
\begin{center}
\begin{tabular}{lccc}
\hline
\hline
Source&z&$\mu_{1/2}$&$M_{I_{C}}$\\
\hline
3C34&0.690&21.7&$-$24.9\\

3C41&0.795&19.4&$-$24.5\\

3C49&0.621&20.5&$-$24.5\\

3C226&0.820&19.7&$-$24.9\\

3C247&0.749&21.2& $-$25.4\\

3C277.2&0.766&19.5&$-$24.4\\

3C337&0.635&19.9&$-$23.7\\

3C340&0.775&18.5&$-$24.4\\

3C352&0.806&21.1&$-$24.7\\

3C441&0.708&20.1&$-$24.7\\
\hline
\hline
\end{tabular}
\caption{The absolute magnitudes and characteristic surface-brightness
of the $10$-object sub-sample ($H_{0}=50,\Omega_{0}=1$)}
\label{smalltab}
\end{center}
\end{table}
The modelling of these objects proceeded in an identical fashion to that of the $z\simeq0.2$ AGN 
sample studied by McLure {\it et al.} (1999a) and Dunlop {\it et al.} (1999). Each object was 
modelled with both a standard Freeman disc template and de Vaucouleurs $r^{1/4}$ profile, with no 
{\it a priori} assumptions being made about position angle, axial ratio or any possible central 
point-source contribution. Prior to the modelling of each source a maximum radius, outside of which 
pixels would be excluded for insufficient signal-to-noise, was determined by an examination of the 
one-dimensional surface-brightness profile. The median value for this outer radius was $4\asec$, 
corresponding to a physical radius of $33$ kpc at the median redshift of $z=0.8$ 
($H_{0}=50,\Omega_{0}=1$). This is reasonably well matched to the outer radius of $12\asec$ 
($50$ kpc) used for each of the radio galaxies in the $z=0.2$ AGN sample, ensuring that the same 
physical region of the galaxy light distribution is under investigation in both redshift regimes. The 
modelling technique strongly preferred a $r^{1/4}$ host galaxy for all $16$ objects analysed.  The 
quality of the fits was good, with no significant residual flux. 

\section{Results}
\label{ress}
\begin{figure}
\setlength{\epsfxsize}{0.3\textwidth}
\centerline{\epsfbox{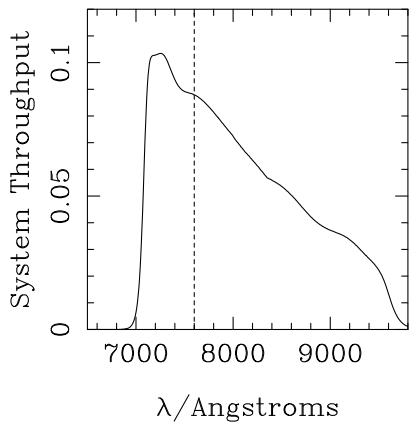}}
\caption{The F814W filter profile including the system response and quantum efficiency of WFPC2. Also shown is the location of the $4000$\AA\, break of an elliptical galaxy spectrum observed at a redshift of $z=0.9$}
\label{814filter}
\end{figure}
In this section the scalelength, absolute luminosity and Kormendy relation results from our two-dimensional modelling of the $z\simeq0.8$ objects are presented.  Due to the significant effect that choice of cosmology can have on angular diameter, cosmological dimming and look-back time over this redshift range, we give these results for a range of possible cosmologies. Four different representative scenarios are considered featuring two values of $H_{0}$ (50,70), together with both open ($\Omega_{0}=0.1$)
 and flat ($\Omega_{0}=1$) geometry.

The results from the modelling of all 16 HST objects are presented in Table \ref{bigtab}.  The main 
conclusions reached from this modelling work are based largely on the results for a 10-object 
sub-sample. The sub-sample consists of the low-redshift end of the BLR sample and is illustrated in 
Fig \ref{subsample}. The remaining 6 objects ($z\ge0.9$) have been excluded from the following 
analysis due to the incursion of the $4000$\AA\, break into the HST F814W filter. The throughput of 
the F814W filter is shown in Fig \ref{814filter} complete with the system response and CCD quantum 
efficiency. As can be seen from this Figure, for objects with redshifts $z\ge0.9$ this filter will 
bridge the $4000$\AA\, break, meaning that it can no longer be assumed that the detected flux has 
originated from the dominant old stellar population. Given that the alignment effect in radio galaxies
is stronger in the rest-frame UV than at longer wavelengths it is to be expected that these objects 
are contaminated by a significant aligned component. The $10$-object sub-sample has the advantage of 
having a tight redshift distribution with a mean of $0.74\pm0.07$ and a median of $0.76$, which is 
well matched by the distribution of the $10$ radio galaxies studied at low-$z$, with its mean redshift of $0.20\pm0.04$ and median of $0.20$. This allows a direct comparison between the two samples without the added complication of allowing for significant evolutionary effects within the samples themselves. 
\subsection{Scalelengths}
\begin{figure}
\setlength{\epsfxsize}{0.3\textwidth}
\centerline{\epsfbox{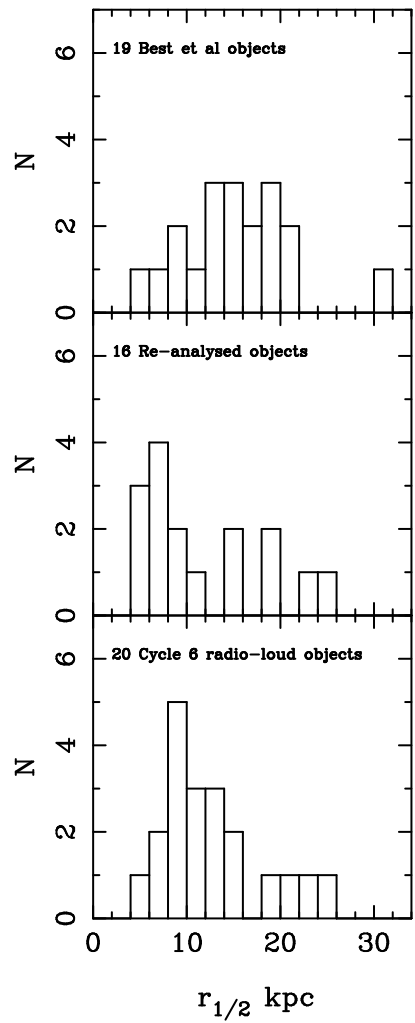}}
\caption{\small{A comparison of the distribution of scalelengths derived from our modelling of the 
$z\simeq0.8$ 3CR galaxies with that obtained by Best {\it et al}. Also shown is the distribution of 
scalelengths obtained for the 10 radio galaxies and 10 radio-loud quasars from our 
HST $z\simeq0.2$ AGN host-galaxy programme (Dunlop et al. 1999; McLure et al. 1999a,b). All three histograms assume $H_{0}=50, \Omega_{0}=1$}}
\label{scales}
\end{figure}
\begin{table}
\begin{center}
\begin{tabular}{cccc}
\hline
\hline
$\Omega_{0}$&$H_{0}$&median $r_{e}$ / kpc&$<r_{e}>$ / kpc\\
\hline
1&50&9.7&$12.6\pm2.3$\\
1&70&8.2&$9.3\pm1.6$\\
0.1&50&11.5&$14.8\pm2.7$\\
0.1&70&8.2&$10.6\pm1.9$\\
\hline
\end{tabular}
\caption{The scalelength results from the two-dimensional modelling of the $z\simeq0.8$ sub-sample. Columns one and two detail the choice of cosmology. Column three gives the median scalelength values for the 10-objects in the sample. Column four gives the corresponding mean values together with the standard error.}
\label{0.8scale}
\end{center}
\end{table}

The results of our determination of galaxy half-light radius ($r_{e}$)
for the 10-objects in the $z\simeq0.8$ sub-sample are presented in
Table \ref{0.8scale}. Two features of this table are immediately
obvious. Firstly, it is clear that regardless of cosmology the derived
half-light radii are remarkably consistent, with both the mean and
median values agreeing to within 4 kpc.  Secondly, the mean scalelength is systematically larger than the median in all four cosmologies, consistent with the suggestion from Fig \ref{scales} that the distribution of scalelengths from the full 16-object sample, and indeed our 20 radio-loud AGN at $z\simeq0.2$, has a substantial tail toward high values.

Given that the scalelength results obtained by BLR are quoted assumming $H_{0}=50, \Omega_{0}=1.0$ it is straightforward to investigate what differences exist between their scalelength determinations and those presented here.  Due to the fact that most of
 the BLR scalelength information is derived from their K--band observations rather than the HST images, it is not the case that they have published a scalelength value for each of the objects re-modelled in this paper. In order to perform a comparison between the two sets of results it has been necessary to simply base the BLR figures on the 8-objects from our 10-object sub-sample for which they have derived a scalelength value.

Assuming $H_{0}=50, \Omega_{0}=1.0$ the mean scalelength of the 10-object sample is $r_{e}=12.61\pm2.26$ kpc with a median of $9.70$ kpc. The corresponding BLR-derived values for $8$ of these $10$ objects are $r_{e}=15.15\pm2.71$ kpc with a median of $14.
95$ kpc. Given the small number statistics that are available, and the inherent difficulty in constraining galaxy scalelengths, the median is the more robust measure of the typical scalelength. Using the median it can be seen from these figures that the one-dimensional analysis technique employed by BLR has systematically overestimated the characteristic size of the 3CR galaxies by $\approx50\%$. 

In order to check that the use of these two cut-down samples was not overtly biasing the results, a comparison was also performed between the full $16$-object sample and the $12$-objects from this sample for which there is also a BLR-derived scalelength figure available. The mean scalelength of the $16$-object sample is $r_{e}=11.48\pm1.59$ kpc with a median of $9.20$ kpc. The corresponding BLR-derived values for $12$ of these objects are $r_{e}=15.66\pm1.91$ kpc with a median of $15.45$ kpc. It can be seen from this that the use of the expanded samples strengthens the conclusion that the BLR scalelengths are overestimated, with the median BLR scalelength being $\approx70\%$ greater than the two-dimensional modelling results presented here. 
\subsection{Absolute Magnitudes}
\label{abmag}
\begin{table}
\begin{center}
\begin{tabular}{ccccc}
\hline
\hline
$\Omega_{0}$ & $H_{0}$ & $z$ &Median  M$_{\rm{I}}$ & $<{\rm M}_{\rm{I}}>$ \\
\hline
1.0&50&0.2&$-$24.52&$-23.45\pm0.16$\\
1.0&50&0.8&$-$24.60&$-24.62\pm0.13$\\
1.0&70&0.2&$-$23.69&$-23.72\pm0.16$\\
1.0&70&0.8&$-$23.87&$-23.89\pm0.13$\\
0.1&50&0.2&$-$24.54&$-24.55\pm0.16$\\
0.1&50&0.8&$-$24.98&$-25.07\pm0.18$\\
0.1&70&0.2&$-$23.80&$-23.81\pm0.16$\\
0.1&70&0.8&$-$24.25&$-24.24\pm0.13$\\
\hline
\hline
\end{tabular}
\end{center}
\caption{The absolute $I$-band magnitudes of the two 10-object radio galaxy sub-samples. Column four 
gives the median figures with column five listing the corresponding mean figures complete 
with standard error.}
\label{mags}
\end{table}
The absolute Cousins $I$-band magnitudes for the $z\simeq0.8$ and $z\simeq0.2$ sub-samples are 
listed in Table \ref{mags} for the four different cosmologies. The values shown are calculated from 
integrating the best-fit de Vaucouleurs profile to infinite radius in order to be consistent with the 
published results for the low-$z$ AGN hosts (McLure {\it et al} 1999a, Dunlop {\it et al} 1999).  
As was pointed out in Section \ref{smalltab} the objects comprising the sub-sample were imaged 
through two separate HST filters; F814W and F785LP. The F814W filter is a member of the standard 
HST filter set and closely mimics the Cousins $I$-band filter.  As a result, it was decided to 
convert all of the integrated apparent magnitudes to their equivalent Cousins $I$-band value.  
The $5$ objects from the sub-sample imaged with the F814W filter had no correction applied to them 
since the similarity of the F814W and Cousins $I$-band filters is such that for elliptical galaxies 
the difference in magnitude is expected to be less than $0.05$ in all cases. This level of photometric accuracy is substantially greater than is possible in the face of the uncertainties in determining the host magnitudes.

The throughput for the F785LP filter is significantly different from that of the Cousins $I$-band 
filter. The conversion from magnitudes obtained through this filter to Cousins $I$-band magnitudes is 
therefore more complicated. The original strategy to overcome this problem was to make use of the one 
object from the $16$-object sample (3C239) for which there are exposures in both filters available 
with comparable signal-to-noise. However, this object has the highest redshift, $z=1.781$, of all of 
the $28$ objects in the BLR sample and subsequently the two images were too faint to get a reliable 
conversion.  In addition to this, 3C239 has a highly distorted morphology in the HST images (BLR 1997), making it unclear whether a conversion factor obtained from this object would be applicable to the 
mostly undistorted objects in the sub-sample. In light of this, the method used to convert the 
F785LP fluxes for each host to the equivalent F814W figure, and hence Cousins I-band magnitude, was 
to predict the count rates $R_{object}$ ($e^{-1}s^{-1}$pixel$^{-1}$) of a source of apparent visual 
magnitude $V$ in the two filters using the filter ratios of Holtzman {\it et al} 1995.

The cosmology-independent k-corrections for each object, produced by the blueward shifting of the 
$I$-band filter along the galaxy spectrum with increasing redshift, were calculated from the 
figures presented for a burst-elliptical galaxy by Rocca-Volmerange \& Guiderdoni (1988). Given that the 
predictions produced by different spectral synthesis codes can differ significantly 
(Charlot {\it et al} 1996) it was considered worthwhile to make an independent check of the validity 
of these k-corrections. Due to the restriction of this analysis to the $10$-object sub-sample, for 
which the images sample galaxy light longward of the $4000$\AA \,break, it is possible to estimate the necessary cosmological k-corrections by modelling the galaxy spectrum as a power-law of the 
form $f_{\nu}\,\alpha\,\nu^{-\alpha}$, where $\alpha$ is the spectral
index. The value of $\alpha$ appropriate for radio galaxies imaged in 
the $I$-band was estimated to be $\alpha\simeq2$ from a 
typical old ($\ge12$Gyr) elliptical galaxy spectrum.  Alternatively,
the mean $I-J$ colour of the sub-sample objects can be used to estimate the
spectral index, a process which also results in a figure of
$\alpha\simeq2$. Both of these estimates are in good agreement with
the value of $\alpha=1.82$ needed to reproduce the k-corrections of
Rocca-Volmerange and Guiderdoni (1988).  The question of whether the 
absolute magnitudes presented in Table \ref{mags} are substantially 
different from the results previously obtained for the $z=0.2$ radio
 galaxies is addressed within Section \ref{comparison}.

\subsection{The Kormendy Relation}
\label{korz}
Given that the hosts of the $10$ radio galaxies within the sub-sample are well fitted by a standard 
de Vaucouleurs galaxy template, it is interesting to see whether the parameters obtained from these 
fits produce a Kormendy relation comparable to that followed by both
low-$z$ inactive ellipticals (and the hosts of the $z=0.2$ AGN, see
Section \ref{comparison}). The $\mu_{1/2}$ values required to construct the Kormendy relation 
have been corrected for the cosmological dimming of surface-brightness according to:
\begin{equation}
\frac{ I_{1} }{ I_{2} }=\frac{ (1+z_{2} )^{3+\alpha}}{ (1+z_{1})^{3+\alpha} }
\end{equation}
\noindent
where a value of $\alpha=1.8$ has been assumed. The resulting
$\mu_{1/2}-r_{1/2}$ relations for the four cosmologies are given in Table \ref{0.8kor}.
\begin{table}
\begin{tabular}{ccc}
\hline
\hline
$\mu_{1/2}=3.51_{\pm0.36}\log r_{1/2}+16.53_{\pm0.38}$&$\Omega_{0}=1.0$&$H_{0}=50$\\
\\
$\mu_{1/2}=3.51_{\pm0.36}\log r_{1/2}+17.05_{\pm0.33}$&$\Omega_{0}=1.0$&$H_{0}=70$\\
\\
$\mu_{1/2}=3.51_{\pm0.38}\log r_{1/2}+16.28_{\pm0.42}$&$\Omega_{0}=0.1$&$H_{0}=50$\\
\\
$\mu_{1/2}=3.51_{\pm0.38}\log r_{1/2}+16.80_{\pm0.37}$&$\Omega_{0}=0.1$&$H_{0}=70$\\
\hline
\hline
\end{tabular}
\caption{The best-fitting Kormendy relations for the 10 $z\simeq0.8$ 3CR galaxies under four different choices of cosmology.}
\label{0.8kor}
\end{table}
\noindent
It is clear from this that the choice of cosmology makes no significant difference to the slope of the Kormendy relation (as expected given the small redshift range of the objects), and that in all cases this slope is consistent with that of $\simeq3$ displayed by inactive low-$z$ ellipticals (Kormendy 1977).

It is worth noting that the $r_{1/2}$ and $\mu_{1/2}$ parameters from the one-dimensional modelling of BLR failed to produce a $\mu_{1/2}-r_{1/2}$ relation which was consistent with the expected slope of $\simeq3$. The modelling results of BLR had the $z\simeq1$ 3CR galaxies lying along a constant luminosity slope of $5$, exactly as is expected when the galaxy luminosities have been well determined but the 
scalelengths have not been constrained (Abraham {\it et al.} 1992). The fact that the relation 
presented above is consistent with the Kormendy relation, within the errors, can therefore be taken 
as further evidence that our two-dimensional modelling has been much more successful in constraining 
the scalelengths of the sub-sample objects. The possibility of using the Kormendy relations derived 
above to test for the effects of passive or dynamical evolution is explored in section \ref{kor}.

\section{Comparison with low-redshift Radio Galaxies}
\label{comparison}
\subsection{Scalelengths}
\begin{table}
\begin{center}
\begin{tabular}{cccc}
\hline
\hline
$\Omega_{0}$&$H_{0}$&median $r_{e}$ / kpc&$<r_{e}>$ / kpc\\
\hline
1&50&11.3&$13.8\pm2.2$\\
1&70&8.1&$9.8\pm1.6$\\
0.1&50&11.9&$14.4\pm2.3$\\
0.1&70&8.5&$10.3\pm1.6$\\
\hline
\end{tabular}
\caption{The scalelength results from the two-dimensional modelling of the $z\simeq0.2$ sub-sample. Columns one and two detail the choice of cosmology. Column three gives the median scalelength values for the 10-objects in the sample. Column four gives the corresponding mean values together with the standard error.}
\label{0.2scale}
\end{center}
\end{table}

The derived median and mean scalelengths for the 10-source
$z\simeq0.2$ radio galaxy sample from our low-$z$ AGN host galaxy
study (McLure {\it et al.} 1999a, Dunlop {\it et al.} 1999) are
presented
 in Table \ref{0.2scale}. A comparison of these results with those presented for the
$z\simeq0.8$ sub-sample in Table \ref{0.8scale} shows the two
groups of galaxies to have very similar 
characteristic scalelengths. The poorest agreement between the two sets of results is for the 
$\Omega=1.0,\, H_{0}=50$ cosmology. However, even in this case the figures are consistent 
to within the errors.  In the other three cosmologies the median and mean scalelengths for the 
two samples are basically identical, differing by $\le0.5$ kpc in all cases. The similarity between 
the scalelength distributions of the two sub-samples is confirmed by an application of the 
Kolmogorov--Smirnov (KS) test. For the two $\Omega_{0}=1.0$ cosmologies the KS test returns a 
probability of 0.68 that the two samples are drawn from the same underlying distribution. 
This conclusion is even stronger in the $\Omega_{0}=0.1$ cosmologies where the KS test returns a 
probability of 0.97 that the two distributions are the same. Therefore,
we find no evidence that the galaxies comprising the $z\simeq0.8$ sub-sample are systematically larger 
than their $z\simeq0.2$ counterparts.

\subsection{Absolute Luminosity}
\label{abmags}
In order to allow a direct comparison of the characteristic
luminosities of the low- and high-redshift 
radio galaxy sub-samples, and thereby look for any evidence of significant merger activity, 
it is first necessary to correct for the different filters and the expected passive evolution of 
the stellar populations.  To achieve this it was decided to convert the $z\simeq0.2$ $R$-band 
magnitudes to their $I$-band equivalents, and then to predict the brightening of the stellar 
population between $z=0.2$ and $z=0.8$ due to passive evolution alone. The present-day colour of an 
elliptical galaxy formed at high redshift ($z\ge3$) has been taken as  $R-I=0.7$ 
(Fukugita {\it et al.} 1995). The corrections to be made for the effects of passive evolution 
have been calculated using the synthetic galaxy spectral models of Jimenez {\it et al.} (1996) 
using our chosen set of four possible cosmologies. Several different galaxy formation redshifts were 
considered to investigate their effects on the resulting correction. A comparison between the 
predicted amount of $I$-band passive evolution between $z=0.2$ and $z=0.8$ and the difference in 
median absolute luminosity of the two samples is presented in Table \ref{passive}.

It is immediately clear from these results that the difference in absolute $I$-band luminosity
 between the $z\simeq0.2$ and $z\simeq0.8$ samples is inconsistent with the amount of passive 
evolution predicted by the stellar synthesis models within the two
 Einstein-de Sitter 
cosmologies. This is still true even in the $\Omega_{0} = 1$ model which allows present-day 
ellipticals to be as old as possible ($z_{for}=10,H_{0}=50$); this model still requires 
2.5 times more magnitudes of passive evolution than is seen in the
 data. 

Adoption of either of the open cosmologies results in a measured
difference in absolute luminosity of $\simeq0.5$ mags between the two
galaxy samples (columns 3 \&\ 4). As can 
be seen from Table \ref{passive}, this luminosity difference is
formally consistent with the expected passive 
evolution in both open cosmologies, for all alternative formation
redshifts. However, it is interesting to note from Table \ref{passive}
that the $H_{0}=70,\Omega_{0}=0.1$ cosmology clearly provides the best
match  between the data and the spectrophotometric model predictions.  In this cosmological 
picture the model predictions are perfectly consistent with the data for all star formation redshifts of 
$z\ge4$, in good agreement with recent discoveries of old stellar population ellipticals at 
redshifts of $z=1.5\rightarrow2$ (Dunlop {\it et al.} 1996, Spinrad {\it et al.} 1997, 
Stiavelli {\it et al.} 1999 ).
\begin{table*}
\begin{tabular}{ccccccc}
\hline
\hline
$\Omega_{0}$ & $H_{0}$ &$\Delta$M$_{med}$&$\Delta$M$_{mean}$ & &$\Delta$M$_{model}$& \\
\hline
 & & & &$z_{for}=10$ & $z_{for}=5$ &$z_{for}=3$\\
\hline
1&50&0.17&0.18$\pm$0.16&0.52&0.60&0.70\\
1&70&0.17&0.18$\pm$0.16&0.63&0.60&0.56\\
0.1&50&0.52&0.44$\pm$0.16&0.26&0.24&0.46\\
0.1&70&0.43&0.45$\pm$0.16&0.38&0.51&0.64\\
\hline
\hline
\end{tabular}
\caption{The results of the tests to determine the influence of different cosmology and formation 
redshift upon the amount of passive evolution expected between $z=0.8$ and $z=0.2$. Columns one and 
two list the cosmological parameters for each test. Column three lists the amount of evolution 
required to reconcile the median of the two absolute magnitude distributions. Column four gives the 
corresponding mean figures complete with standard errors. Columns 5-7 list the amount of passive 
evolution between $z=0.2$ and $z=0.8$ predicted by our spectrophotometric modelling for 
three galaxy formation redshifts. (The apparently anomalous trend displayed by the $\Omega_{0}=1, H_{0}=70, z_{for}=3$ model is due to the fact that even at $z=0.2$ a galaxy is only 6 Gyr old, and remains relatively bright at I.)}
\label{passive}
\end{table*}
\subsection{Kormendy Relation}
\label{kor}
\begin{figure*}
\setlength{\epsfxsize}{0.7\textwidth}
\centerline{\epsfbox{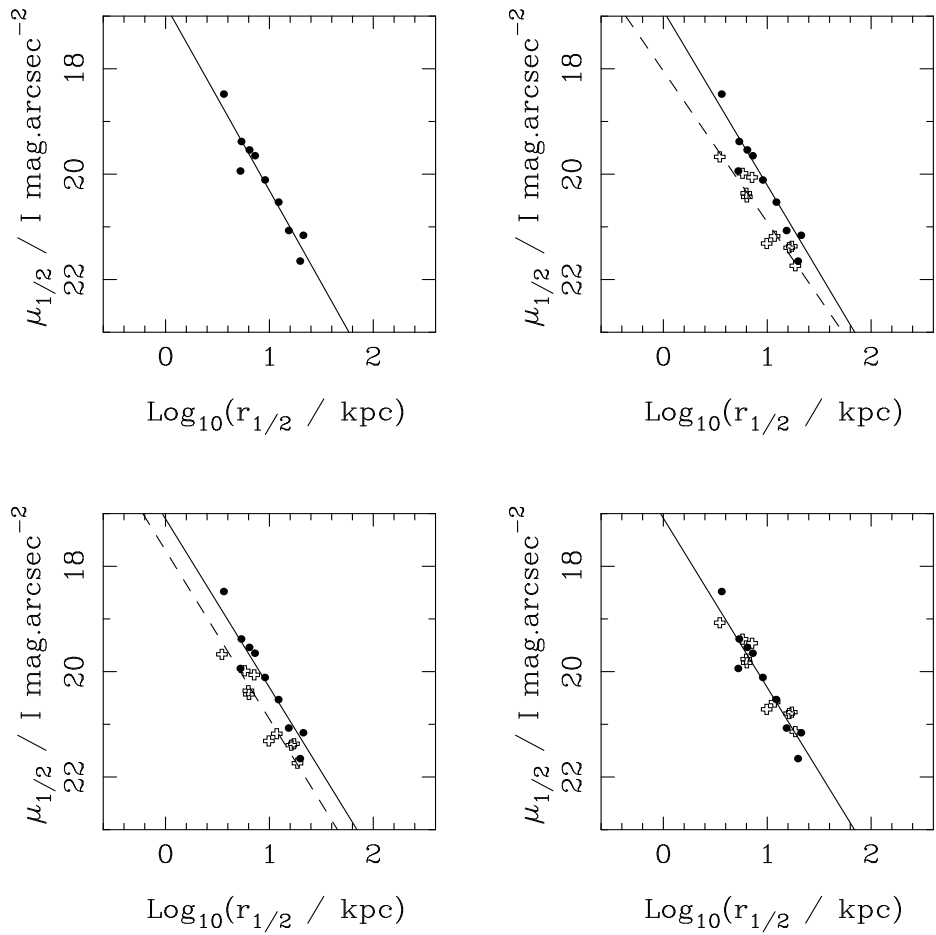}}
\caption{Shown in the top-left panel is the Kormendy relation followed by the $z\simeq0.8$ sub-sample 
which has a best-fit slope 0f 3.5. In the top-right panel the $z\simeq0.2$ radio galaxies (crosses) 
have been added along with their best-fit relation of slope 2.9 (dashed line). In the bottom-left 
panel both galaxy sub-samples are shown with the best-fit relation forced to have a intermediate 
slope of 3.20. The bottom right figure shows the best-fit Kormendy relation (slope=3.21) produced by 
brightening the surface-brightness of the $z\simeq0.2$ galaxies by 0.6 magnitudes.}
\label{thekors}
\end{figure*}
\begin{table}
\begin{tabular}{ccc}
\hline
\hline
$\mu_{1/2}=2.86_{\pm0.34}\log r_{1/2}+17.66_{\pm0.38}$&$ \Omega_{0}=1.0$&$H_{0}=50$\\
\\
$\mu_{1/2}=2.86_{\pm0.34}\log r_{1/2}+18.08_{\pm0.33}$&$ \Omega_{0}=1.0$&$H_{0}=70$\\
\\
$\mu_{1/2}=2.85_{\pm0.34}\log r_{1/2}+17.61_{\pm0.42}$&$ \Omega_{0}=0.1$&$H_{0}=50$\\
\\
$\mu_{1/2}=2.85_{\pm0.34}\log r_{1/2}+18.03_{\pm0.37}$&$ \Omega_{0}=0.1$&$H_{0}=70$\\
\hline
\hline
\end{tabular}
\caption{The best-fitting Kormendy relations for the 10 $z\simeq0.2$ radio galaxies under four 
different choices of cosmology.}
\label{0.2kor}
\end{table}

\begin{table}
\begin{center}
\begin{tabular}{ccccc}
\hline
\hline
$\Omega_{0}$ & $H_{0}$ &$\Delta M (-1\sigma)$&$\Delta M$&
$\Delta M(+1\sigma)$ \\
\hline
1.0&50&-0.24&0.43&0.56\\
\hline
1.0&70&-0.24&0.44&0.58\\
\hline
0.1&50&\,\,0.18&0.60&0.84\\
\hline
0.1&70&\,\,0.19&0.60&0.84\\
\hline
\hline
\end{tabular}
\end{center}
\caption{The amount of luminosity evolution predicted by forcing both galaxy 
sub-samples to lie on a Kormendy relation with slope 3.2. Column four lists the implied vertical shift 
required to over-lay the Kormendy relations of the two samples using the best-fit scalelengths of Table \ref{bigtab} converted to the appropriate cosmology. Columns 3 and 5 list the vertical shift implied by shifting the best-fit scalelengths left($-1\sigma$) and right($+1\sigma$) until the scalelength distributions of the two sub-samples differ at the $1\sigma$ level. }
\label{shifts}
\end{table}

If the two samples of radio galaxies can truly be linked by a single population of passively 
evolving ellipticals it is to be expected that they should follow
Kormendy relations which are identical except for a simple vertical shift in surface-brightness. After making the appropriate surface-brightness 
corrections (Section \ref{korz}) and $R \rightarrow I$-band filter transformation (Section \ref{abmags}) the 
least-squares fit to the Kormendy relations formed by the $z\simeq0.2$ radio galaxies are shown in 
Table \ref{0.2kor}.

It can be seen from this that, as for the $z\simeq0.8$ galaxies, the choice of cosmology makes little 
difference to the slope of the Kormendy relation, although the
normalization does change significantly. It is also apparent that the
low-$z$ objects appear to follow a substantially flatter relation,
with a slope of 2.9 instead of 3.5, although clearly both values are
consistent with the expected slope $\simeq3$ (due to the fairly
substantial formal error in the fitted slope, which is predominantly a
result of the small sample size and lack of dynamic range in radius).
To facilitate a fair comparison of the high- and low-$z$ radio galaxy 
Kormendy relations, the least-squares fitting was repeated with 
an enforced intermediate
slope of 3.20 . If it is indeed the case that these two galaxy
populations can be linked by passive evolution alone then the vertical
shift required to reconcile the Kormendy relations should be in good
agreement with both that required to match the absolute magnitudes,
and the passive evolution predictions of the stellar synthesis
modelling. To explore what range of vertical magnitude shifts are
allowed by the data, we performed the least-squares fitting (with
fixed slope=3.20) not only with the best estimate of $r_{e}$ for each
galaxy, but with the
scalelengths of the $z\simeq0.8$ sub-sample shortened and lengthened
such that the KS test showed its scalelength distribution to differ from that
of the $z\simeq0.2$ sub-sample at the $1\sigma$
level. The results of this process are presented in Table
\ref{shifts}. A comparison of the figures from Table \ref{shifts} with
those of Table \ref{passive} shows that in the $\Omega_{0}=1.0$ models
considered here it is only possible to reconcile the luminosity
evolution predicted by the Kormendy relations with that predicted by
the spectral modelling and the absolute magnitude distributions,
separately.
 In these cosmologies it is impossible to force both alternative
measures of the amount of luminosity evolution, and the predictions of
the spectral modelling, into agreement. In
contrast, the $\pm1\sigma$ predictions from the Kormendy relations in
the $\Omega_{0}=0.1$ cosmologies comfortably bracket the offsets given
in Table \ref{passive}, thus leaving both alternative measures of the
luminosity evolution in excellent agreement.

In theory, the Kormendy relations for the low and high-z sub-samples
presented here offer an opportunity to constrain both cosmology and
the prevalence of merger activity. However, in practise the effects of
these are very closely coupled. For example, our results can be
reproduced either by pure passive evolution since $z\simeq0.8$ in an
open Universe, or by modest growth ($\simeq20\%$ growth in scalelength
\&\ luminosity) since $z\simeq0.8$ in an Einstein-de Sitter
Universe. Despite this inherent degeneracy the Kormendy relations can
firmly exclude very strong growth ($\ge 50 \%\ $) from mergers in the
redshift range $0.2<z<0.8$.
 
\section{Conclusion}

The results from a thorough re-examination of the BLR HST images of a sample of $z\sim1$ 
3CR radio galaxies have been presented.  It has been shown that, in terms of scalelength, absolute magnitudes and Kormendy relation there are no significant 
differences between $z\sim0.8$ and $z\sim0.2$ 3CR radio galaxies.  The two populations appear to be 
fully consistent with being comprised of old stellar populations formed at high redshift and 
evolving passively thereafter.  

It is obviously true that the fact that both the high and low-redshift radio galaxies have absolute 
luminosities consistent with pure passive evolution can easily be reconciled with the involvement 
of some dynamic evolution. In the dynamical model this simply requires
that the proto-galactic clumps which merge to produce the final
galaxies where formed reasonably coevally. The results presented here
do not then require that these radio galaxies must have formed in a
monolithic collapse at a single redshift. However, they do suggest
strongly that the vast majority of merger activity within this
population of massive ellipticals must have been completed before
$z\simeq1$. Although semi-analytical galaxy formation models in a 
$\Omega_{0}=0.3, \Lambda=0.7$ cosmology predict as much as 70\% of
present day ellipticals to already be in place by $z\simeq1$ (Kauffman 
\& Charlot 1999), it is worth remembering that the same is not true of
BCGs, which are still predicted to grow by a factor of $2\rightarrow4$
between $z=1$ and the present day (Arag\'{o}n-Salamanca {\it et al.}
1998).

The crucial question still to be answered is whether or not it can be
proven that a significant difference in cluster environment does exist
between the two populations. If it can be shown that this is
definitely the case then the argument forwarded by BLR that we are
observing the effects of dynamical evolution producing the
characteristic mass required for powerful radio emission at the two
different epochs remains tenable; the correlation between
luminosity and scalelength would actually predict the $z\simeq 0.8$
and $z\simeq 0.2$ sub-samples to have similar scalelength
distributions. However, as was discussed in Section \ref{background},
while there have been numerous surveys carried out recently tackling the environments of high-$z$ radio sources, the work on the low-$z$ environments looks to be subject to possible systematic error. This suspicion is further strengthened by our new deep R--band HST images of 
the 10 $z\simeq0.2$ radio galaxy sub-sample (McLure {\it et al.} 1999a, Dunlop {\it et al.} 2000). These images show numbers of apparent companion objects which appear consistent with Abell classes $0\rightarrow2$. This is in good agreement with our recent 
work to model the first to fourth-ranked brightest cluster galaxies (BCG) in clusters spanning Abell classes $1\rightarrow4$, imaged with HST (Dunlop {\it et al.} 1999). This work has shown that the scalelengths of these objects are not significantly different from the radio galaxy scalelength results presented here. 

In an attempt to disentangle the radio galaxy environment problem, more near-infrared observations of the low-$z$ radio galaxies and quasars from our HST programme are planned, to compliment our existing R and B--band images of this sample. This combined 
dataset will allow a fully qualitative investigation of the cluster environments of the low-$z$ radio galaxies with which to compare with the existing high-$z$ environmental data.

If it transpires that there is no significant difference in 3CR
 environments at high and low-$z$, then a much simpler picture of
 powerful radio galaxy evolution emerges. Combined with the
 scalelength evidence presented here, this would suggest that the host
 galaxies of powerful radio sources are basically the same sort of
 objects at {\it all} redshifts from $z=0\rightarrow1$. In this picture the fall-off in 3CR radio power over this redshift range would simply be due to the progressively smaller amount of gas available to feed the central engine. 

\section{Acknowledgements}

We thank the referee Philip Best for comments which undoubtedly
improved a number of aspects of this paper. Based on observations with
the NASA/ESA Hubble Space Telescope, obtained
at the Space Telescope Science Institute, which is operated by the
Association of Universities for Research in Astronomy, Inc. under NASA
contract No. NAS5-26555.
This research has made use of the NASA/IPAC Extragalactic Database (NED)
which is operated by the Jet Propulsion Laboratory, California Institute
of Technology, under contract with the National Aeronautics and Space
Administration.
MJK acknowledges the award of a PPARC PDRA, and also acknowledges
support for this work provided by NASA through grant numbers O0548 
and O0573 from the Space Telescope Science Institute, which
is operated by AURA, Inc., under NASA contract NAS5-26555. RJM
acknowledges a PPARC studentship.

\section{References}
\noindent
Abraham R.G., Crawford C.S., McHardy I.M., 1992, ApJ, 401, 474\\
Arag\'{o}n-Salamanca A., Baugh C.M., Kauffmann G., 1998, MNRAS, 292, 427\\
Best P.N., Longair M.S., R\"{o}ttgering H.J.A., 1997, MNRAS, 292, 758\\
Best P.N., Longair M.S., R\"{o}ttgering H.J.A., 1998, MNRAS, 295, 549\\
Bruzual G., Charlot S., 1993, ApJ, 405, 538\\
Charlot, S., Worthey G., Bressan. A., 1996, ApJ, 457, 43\\
Collins C.A., Mann R.G., 1998, MNRAS, 297, 128\\
Dickinson M., 1997, in Tanvir N.R., Aragon-Salamanca A., Wall J.V., eds, HST and the high redshift Universe. Singapore: World Scientific, p.207\\
Dunlop J.S., Peacock J.A., 1993, MNRAS, 262, 936\\
Dunlop J.S., Taylor G.L., Hughes D.H., Robson E.I., 1993, MNRAS, 264, 455\\
Dunlop J.S., et al., 1996, Nature, 381, 581\\
Dunlop J.S., et al., 2000, MNRAS, in preparation\\
Eales S., Rawlings S., Law-Green D., Cotter G., Lacy M., 1997, MNRAS, 291, 593\\
Fukugita M., Shimasaku K., Ichikawa T., 1995, PASP, 107, 945\\
Hill G.J., Lilly S.J., 1991, ApJ, 367, 1\\
Holtzman J.A., et al, 1995, PASP, 107, 1065\\
Jimenez R., MacDonald J., 1996, MNRAS, 283, 721\\
Kauffmann G., Charlot S., 1999, astro-ph/9810031\\
Kormendy J., 1977, ApJ, 217, 406\\
Krist J., 1995, Astronomical Data Analysis and Systems IV, ASP Conference Series, Vol 77, R.A. Shaw, H.E. Payne, and J.J.E. Hayes, eds, 349\\
Laing R.A., Riley J.M., Longair M.S., 1983, MNRAS, 204, 151\\
Leyshon G., Eales S.A., 1998, MNRAS, 295,10L\\
Lilly S.J., Longair M.S., 1984, MNRAS, 211, 833\\
Lilly S.J., Prestage R.M., 1987, MNRAS, 225, 531\\
Oke J.B., Gunn J.E., 1983, ApJ, 226, 713\\
Madore B.F., et al., 1999, ApJ, 515, 29\\
McCarthy P.J., van Breugel W.J.M., Spinrad H., Djorgovski S., 1987, ApJ, 321, L29\\
McLure R.J., et al., 1999a, MNRAS, 308, 377\\
McLure R.J., et al., 1999b, MNRAS, submitted\\
Perlmutter S., et al., 1999, ApJ, 517, 565\\
Prestage R.M., Peacock J.A., 1988, MNRAS, 230, 131\\
Rocca-Volmerange B., Guiderdoni B., 1988, A\&AS, 75, 93\\
Schombert J.M., 1987, ApJS, 64, 643\\	
Spinrad H., et al., 1997, ApJ, 484, 581\\
Stiavelli M., et al., 1999, A\&A, 343, L25\\
Taylor G.T., Dunlop J.S., Hughes D.H., Robson E.I., 1996, MNRAS, 283, 930\\
\end{document}